\documentclass[11pt,twoside]{article}
\usepackage{asp2004}
\usepackage{psfig}
\usepackage{epsf}
\usepackage{graphics}
\usepackage{lscape}
\def\edcomment#1{\iffalse\marginpar{\raggedright\sl#1\/}\else\relax\fi}
\reversemarginpar

\begin{document}
\title{Polarimetry of powerful radio galaxies from $z=0$ to $z=4$}
\author{Clive Tadhunter}
\affil{Department of Physics \& Astronomy, University of Sheffield, Sheffield S3
7RH, UK}

\begin{abstract}
The advent of sensitive polarimeters on large
telescopes has led to a revolution in our understanding of active galaxies over the last 20
years. In the case of powerful radio galaxies the deep polarimetric measurements made
possible by the new technology have: (a) provided the most direct evidence to support the
unified schemes for powerful radio galaxies; (b) enhanced our understanding of the
colours and morphologies of radio source host galaxies at all redshifts; and (c) provided
key information about the kinematics and geometries of the scattering regions on a sub-kpc scale.
\end{abstract}
\thispagestyle{plain}

\section{Introduction}
Active galactic nuclei (AGN) are luminous, radiate their continuum light anisotropically, and are
embedded in a rich ISM that contains plentiful scattering particles such as electrons and
dust grains. It is not surprising, therefore, that polarimetric studies of the scattered light
component provide a key tool for investigating the geometry, kinematics and host
galaxy properties of active galaxies. In this review I will concentrate on polarimetry
of powerful radio galaxies, which form a well-defined class of active galaxies, selected 
on the basis of their extended radio emission. Apart from clean selection criteria, radio galaxies 
have the additional advantage that the directions of their radio jets provide useful geometrical
information.

Polarimetry is a ``photon hungry'' subject, but radio galaxies are relatively faint and
distant. Therefore the detailed polarimetric study of powerful radio galaxies at optical
wavelengths only began in earnest in the early 1980s when polarimeters began to be 
used with sensitive electronic detectors on large telescopes.

The pioneering work was done by Antonucci (1982, 1984) who reported
spectropolarimetric observations of 31 nearby 3C radio galaxies ($z < 0.26$) made with the IDS on
the Lick telescope. Although his survey was 
limited in its sensitivity, Antonucci's results prefigure much of the modern polarimetric
study of radio galaxies. For example, Antonucci was the first to deduce that most of the
optical polarization properties of nearby radio galaxies can be explained in terms of
scattered AGN light, with the variety in the alignments of the polarization E-vectors
accounted for by different configurations of the scattering medium. Antonucci was also the 
first to propose the presence of an optically thick central obscuring region in radio
galaxies, requiring scattering at the poles of a torus structure to explain the perpendicular 
alignments of the optical polarization vectors relative to the radio axes in some objects.

Following Antonucci's early work, the use of sensitive CCD-based waveplate
polarimeters on 4m telescopes in the late 1980s allowed accurate imaging polarimetry 
measurements to be made of more distant radio galaxies out to a redshift of $z\sim1$.  Then,
the subsequent introduction of 8m class telescopes both further extended the redshift 
range available for polarimetric investigations of distant radio galaxies (up to $z\sim4$), and
also facilitated deeper spectropolarimetric studies of nearby sources. Finally, since 1994 the
polarimeters on the post-COSTAR HST have allowed high spatial resolution imaging
polarimetry observations to be made of the inner structures on a sub-kpc scale at both
optical and dust-penetrating near-IR wavelengths.

In the following I will demonstrate that the high sensitivity polarization measurements of the
scatterd light component
allowed by these technological innovations have led to major advances in several key
areas of radio galaxies research. A full list of references for optical/UV polarization
studies of the scattered light in powerful radio galaxies is presented in the Table 1 of
the Appendix.

\section{Polarimetric tests of the unified schemes} 

The unified schemes for powerful radio sources provide a useful framework for
understanding the relationships between the various sub-classes of radio loud AGN. In
the strongest versions of such schemes it is proposed that radio galaxies and radio-loud
quasars are the same thing viewed from different directions, with the quasar light blocked
off from our direct view in the radio galaxies by central obscuring tori (e.g. Barthel
1989). Until the early 1990s attempts to test such schemes were indirect and involved the
statistical comparisons between the radio and optical properties of samples of radio
galaxies and radio-loud quasars (e.g. Barthel 1989, Jackson \& Browne 1990). 
%For example, both the distributions of radio source
%diameter and the radio de-polarization asymmetries for the two classes were found to be
%consistent with the unified schemes. 
However, such statistical comparisons are not capable of providing a true test by
determining whether or not {\it individual} radio galaxies contain quasar nuclei hidden in their cores.

Polarimetric observations have the potential to provide a more direct test of the unified
schemes. If the quasar is luminous enough and there is sufficient scattering ISM in the
quasar illumination cones, a measurable amount of the optical/UV light of the hidden
quasar nucleus will be scattered into our line of sight. The scattered component will be highly
polarized. Thus measurements in polarized light enhance the contrast of the scattered
component relative to the (unpolarized) direct emission from the stars and gas in the host
galaxies. In this way it is possible to detect the scattered quasar emission polarimetrically,
even if it is not readily apparent in the straight intensity images and spectra.

In fact, the first polarimetric detection of this scattered quasar component came well
before the unified schemes gained prominence: in his pioneering work Antonucci (1984) 
detected a broad quasar-like H$\alpha$ line in the
polarized intensity spectrum of the N-galaxy 3C234.

The detailed study of the
scattered quasar component had to await the introduction of sensitive CCD-based
waveplate polarimeters. First, imaging polarimetry observations using 4m class
telescopes demonstrated that the nuclei of some low redshift radio galaxies are
surrounded by spatially resolved optical reflection nebulae, with the polarization
E-vectors aligned perpendicular to the radius vectors from the nuclei, as expected for
scattered nuclear light. Examples include PKS2152-69 (di Serego Alighieri et al. 1988) and 
Cygnus A (Tadhunter et al. 1990). Subsequently,
several more reflection nebulae were  detected in radio galaxies across a wide range of
redshift (e.g. Tadhunter et al. 1992, Cimatti et al. 1993, Draper et al. 1993). 
Most recently, deep spectropolarimetry observations with 8m telescopes have 
detected the broad permitted lines (e.g. H$\alpha$, MgII) characteristic of scattered quasars in the polarized
intensity spectra of nearby radio galaxies (Ogle et al. 1997, Cohen et
al. 1999). These latter observations represent the best,
most direct evidence that at least {\it some} powerful radio galaxies contain luminous hidden
quasar nuclei.

However, several factors can make the scattered quasar component difficult to
detect. These include (depending on the object): a lack of a sufficient scattering medium; 
relatively low luminosity
quasar nuclei; dilution by unpolarized starlight of the host galaxy; dilution
by unpolarized AGN-related components (see below); and the geometrical dilution caused
by integrating the polarization vectors across the broad illumination cones. Therefore, it is 
doubtful that polarimetric observations ever will provide an answer to the question
of whether {\it all} radio galaxies contain hidden quasar nuclei, as suggested by the strongest
versions of the unified schemes.

\section{The UV excess in powerful radio galaxies: starbursts or AGN pollution?}

One of the major motivations for studying powerful radio galaxies is that they represent
potentially important probes of the evolution of giant elliptical galaxies in the high
redshift universe.  However, if we are to use radio galaxies in this way it is essential to
understand their continuum spectral energy distributions (SEDs). In particular, it is crucial to
distinguish the continuum components that are specifically related to the star formation in the host
galaxies, from those that are a direct consequence the AGN activity.

In the early 1980s photometric studies of the optical-infrared colours of the radio galaxies
at $z >0.5$ demonstrated that such objects show significant UV excesses compared with
normal, passively-evolving early-type galaxies (e.g. Lilly \& Longair 1984). At the same time 
imaging studies revealed that
the excess UV light is distributed in elongated, multi-modal structures that are quite
unlike those of elliptical galaxies in the low redshift universe (Lilly, Longair \& McClean 1983). 
Both the colours and UV
structures of the high-z radio galaxies were initially interpreted in terms of starbursts
associated with the evolution of the host galaxies. However, with the discovery that the
UV structures are closely aligned with the axes of the large-scale radio structures ­-- the
so-called ``alignment effect'' (McCarthy et al. 1987, Chambers et al. 1988) ­-- it became clear 
that the extended UV emission might be
related more to the effects of the activity than the evolution of the host galaxies. Indeed, it
was proposed that aligned UV structures represent the giant reflection nebulae
illuminated by the hidden quasar nuclei (Tadhunter et al. 1988, Fabian 1989). 
In this case the UV light should be highly
polarized.

\begin{figure}[!ht]
\plotone{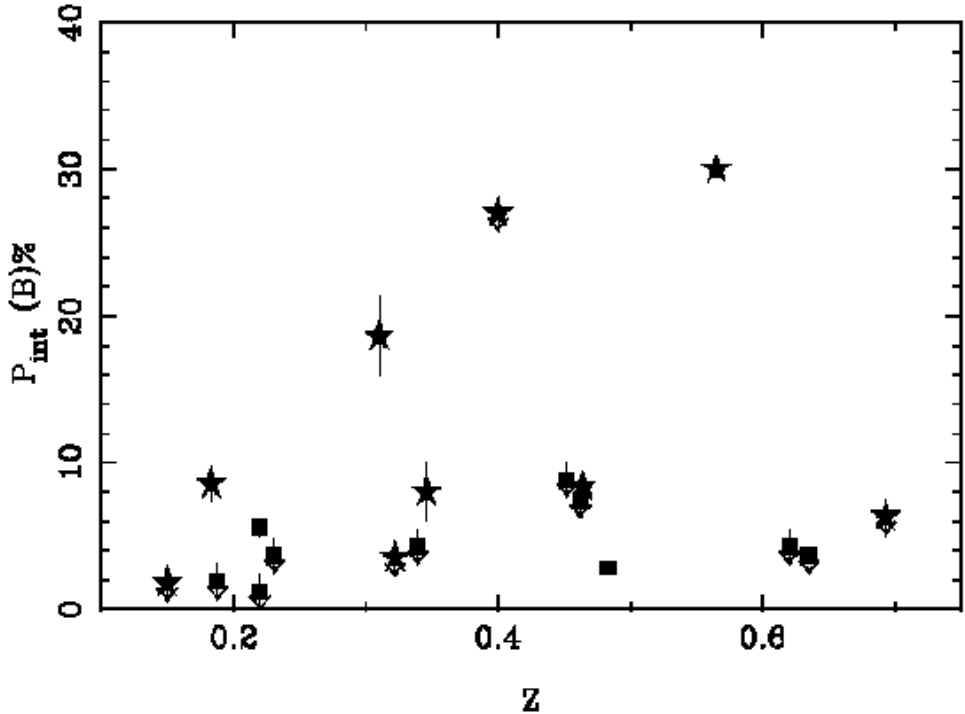}
\caption{B-band polarization plotted as a function of redshift for the 2Jy sample (see Tadhunter et al. 2002
for details). The measured polarization values have been corrected for the dilution by the light of
the old stellar populations and the nebular continuum. They have also been corrected for
Ricean bias. Stars indicate narrow line radio galaxies, whereas squares indicate broad line
radio galaxies. Upper limits are indicated by arrow symbols.}
\end{figure}

The first attempts to test the reflection nebula idea using CCD-based polarimeters on 4m
telescopes proved successful (di Serego Alighieri et al. 1989, Januzzi \& Elston 1991, Tadhunter et al. 1992). 
High degrees of linear polarization were detected in the UV
continua of several high redshift radio galaxies. Moreover, the polarization E-vectors 
were found to be aligned perpendicular to axes of the large-scale radio and UV structures,
as expected for anisotropic illumination by a central point source. With the advent of 8m 
telescopes it became possible to map the polarization of the high-z objects in detail,
perform spectropolarimetry, and extend the polarimetric studies out to higher redshifts (see Table 1
for a full list of references). 
The main general conclusions of these studies are as follows.
\begin{itemize}
\item {\bf Incidence of scattered UV emission}. The extended UV emission is dominated by 
scattered quasar light in some, but not all,
high redshift radio galaxies.

\item {\bf Scales of the refelection nebulae.} The reflection nebulae can be galaxy scale, 
extending 10s of kpc from the nuclei of the 
host galaxies in some objects.

\item {\bf Nature of the scattering medium.} The dominant scattering particles are likely to be dust grains; the
Thompson depth for electron scattering is insufficient to produce the luminosity of the scattered
light in most objects.

\item {\bf Distribution of scattering particles.} The scattering is approximately grey, and the 
scattering dust is likely to be 
distributed in small optically thick clouds (Vernet et al. 2001, Kishimoto et al. 2001).
\end{itemize}

Although these results help to explain the alignment effect in high redshift radio galaxies, 
and lend strong support to the orientation-based unified schemes (see above), they also
suggest that much of the UV emission in these objects represents ``AGN pollution''. 

This raises the issue of the extent to which the scattered quasar light dominates the excess
UV emission in the general population of powerful radio galaxies. Accurate polarimetric 
observations of distant radio galaxies are challenging. Therefore, the early studies were
biased towards the brightest, most spectacular objects in any redshift range and do not 
necessarily provide an accurate indication of the importance of the scattered quasar
component. In order to avoid this selection effect we have carried out a combined 
spectroscopic and imaging polarimetric survey of a complete, optically unbiased sample
of 19 southern 2Jy radio galaxies with redshifts in the range $0.15 < z < 0.7$ (Tadhunter et al. 2002). 
All of the 
objects in this sample show evidence for UV continuum excesses, and by making the
polarimetric observations in the B-band we have  sampled the rest-frame UV in all the objects in 
the sample. The results are shown in Figure 1, where the polarization measurements
have been corrected for dilution by the light of the old stellar populations in the host 
galaxies and nebular continuum. Taking into account the geometrical dilution caused by 
integrating the
polarization over the broad quasar illumination cones, we would expect to measure
polarization at the $>$10\% level if the UV excess is dominated by scattered light (Manzini \& 
di Serego Alighieri 1996). While a
few objects in our sample are polarized at such a high level, most objects are polarized at
a lower level, suggesting that their excess UV emission is not dominated by
scattered quasar light.

What other components might contribute to the UV excess? Careful modelling of the 
polarimetric results in combination with the emission line and continuum spectra for the
2Jy sample has revealed a complex picture. In fact, we find that the following four components
contribute to the UV excess to varying degrees.
\begin{itemize}
\item {\bf Scattered quasar light.} This is detected in 32\% of objects, and dominates in $\sim$15\% of
objects.

\item {\bf Nebular continuum.} This is significant in all objects with strong emission lines,
contributing 3 ­- 40\% of the excess UV emission (see Dickson et al. 1995).

\item {\bf Direct AGN light.} The unpolarized continua of low-luminosity, or partially obscured,
broad line nuclei make a substantial contribution to the UV excess in 40\% of the
objects in our sample.

\item {\bf The light of young stellar populations (YSP).} This makes a significant contribution to the
UV excess in 30 ­- 50\% of the objects in our sample.

\end{itemize}

Figure 2 illustrates the various components that contribute to the UV excess.
The complexity revealed by our polarimetric/spectroscopic survey of the 2Jy sample
is also apparent in high resolution HST images of powerful radio galaxies such as Cygnus A (Jackson, Tadhunter
\& Sparks 1998). 
However, in spite
of the complexity, our results demonstrate that it is possible to detect the young stellar
populations and investigate the evolution of host galaxies, provided that high quality
spectroscopic and polarimetric data are available.

An important lesson from these studies is that it is dangerous to interpret the colours and SEDs of distant active
galaxies
solely in terms of the
properties of the stellar populations in the host galaxies without first quantifying the degree of AGN pollution.

\begin{figure}[!t]
\plotone{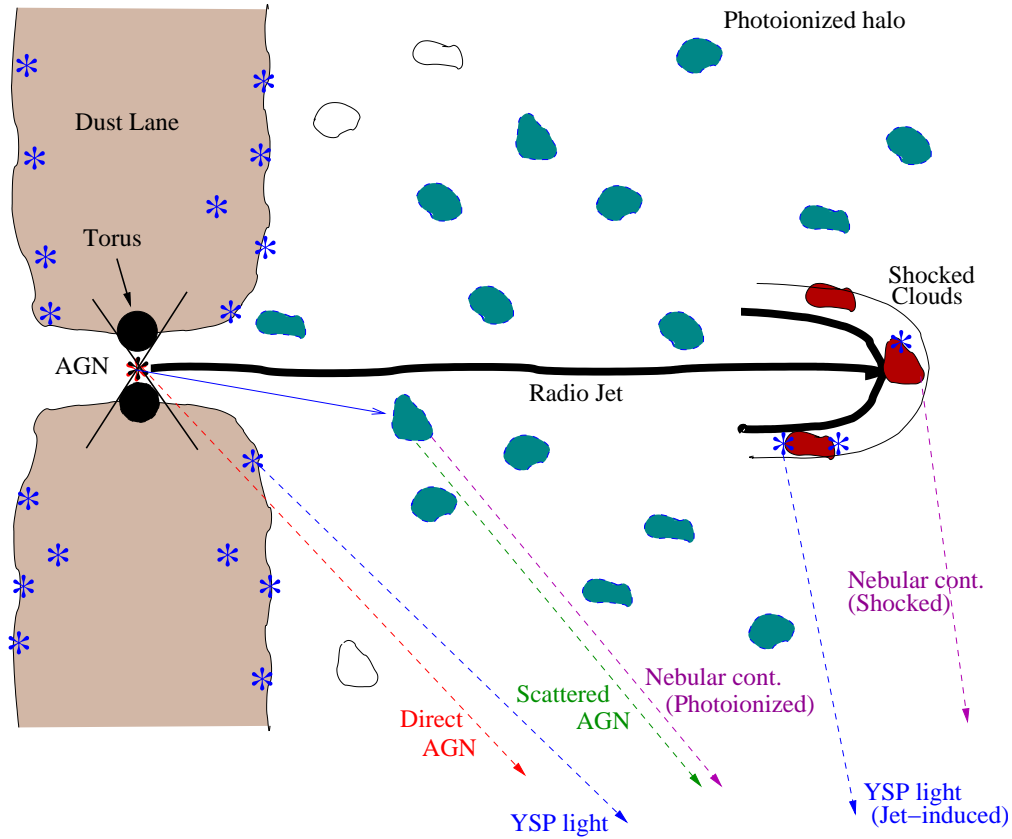}
\vglue 0.5cm\noindent
\caption{Components contributing to the UV excess in powerful radio galaxies.
Components include: scattered AGN light, nebular continuum from the warm
emission line gas (both shocked and photoionized), direct AGN light from
partially obscured AGN, and the light of the young stellar populations (YSP) in
the host galaxies.}
\end{figure}

\section{Investigating the structures and kinematics on a sub-kpc scale}
  
Recently, attention has turned to using polarimetric observations to investigate the
structures and kinematics on the sub-kpc scale --- an important scale for investigating the impact of the
activity on the host galaxies, as well as the structure of the emitting regions in the
immediate vicinity of the super-massive black holes. The contributions by J. Smith and
M. Kishimoto at this conference demonstrate the potential of spectropolarimetric observations
for studying the BLR and inner accretion disks of active galaxies (see Smith
et al. 2004, Kishimoto et al. 2003). I will give two further
examples of the use of polarimetry as a tool for probing the central regions of radio-loud
active galaxies.

\subsection{Near-IR polarimetry observations of Cygnus A}

Near-IR observations can provide further tests of the unified schemes
and allow investigations of the radiation field anisotropy and continuum emission
mechanisms at longer wavelengths than the optical observations described above. The
near-IR observations also have the advantage that, since they are less sensitive to dust
obscuration, they have the potential to probe regions closer to the central AGN.

Some powerful radio galaxies have near-IR core
sources that are unresolved in ground-based observations (e.g. Simpson et al. 1995). Although these compact core
sources have been interpreted in terms of the quasar light ``shining through''  the torus at
less heavily obscured near-IR wavelengths, the extinctions deduced from the near-IR
colours and fluxes are significantly lower than the those deduced from X-ray
measurements of the absorbing columns (see Tadhunter et al. 1999).

In order to investigate the true nature of the compact near-IR core sources we have used
the capabilities of NICMOS on the HST to make high spatial resolution imaging
polarimetry observations of the archetypal radio galaxy Cygnus A at 2.0$\mu$m (Tadhunter et al. 1999, 2000). 
The results
are shown in Figure 3. The intensity images reveal a compact core source that is
unresolved at the $<$0.2 arcsec resolution of the observations, but this core source is highly polarized at the
$>$28\% level, with the polarization E-vector aligned perpendicular to the radio axis. It is
difficult to reconcile the near-IR colours and fluxes with the large extinction required to
produce the polarization  by dichroic extinction. Therefore it is likely that the core source
does not represent transmitted quasar light, but rather quasar light scattered by the far
wall of the torus or by a polar scattering region. This helps to explain the discrepancy
between near-IR and X-ray based estimates for the torus extinction in this source.

The near-IR imaging observations also reveal an extended, edge-brightened bi-cone
structure that is likely to be due to a combination of scattered quasar light and redshifted
Pa$\alpha$ emission. In contrast to the optical
polarization map (Ogle et al. 1997), the optical emission line cones (Jackson et al. 1998), 
and the near-IR 2.0$\mu$m intensity image (Figure 3), which all show 
reflectional symmetry about the radio axis,
the near-IR polarization is concentrated along the NW-SE wall of the bicone. This
suggests that, while the short wavelength X-ray to optical radiation is isotropic within the
cones defined by the torus, the near-IR radiation is {\it intrinsically anisotropic} within the
cones. The most plausible explanation for such intrinsic anisotropy is that the near-IR 
continuum associated with the AGN is emitted by the warped outer part of the accretion
disk; the NW-SE wall sees a larger surface area of the near-IR emitting region than the 
SW-NE wall.  Indeed, such warping 
is predicted by models for radiation induced instabilities in accretion disks (e.g. Pringle 1997).

These observations of Cygnus A demonstrate the potential of high resolution near-IR
polarimetry observations for investigating the radiation field anisotropy in structures
ranging from the sub-pc scale outer accretion disk surrounding the central AGN, to the
kpc-scale reflection nebula associated with the dust lane.

\begin{figure}[!t]
\plotone{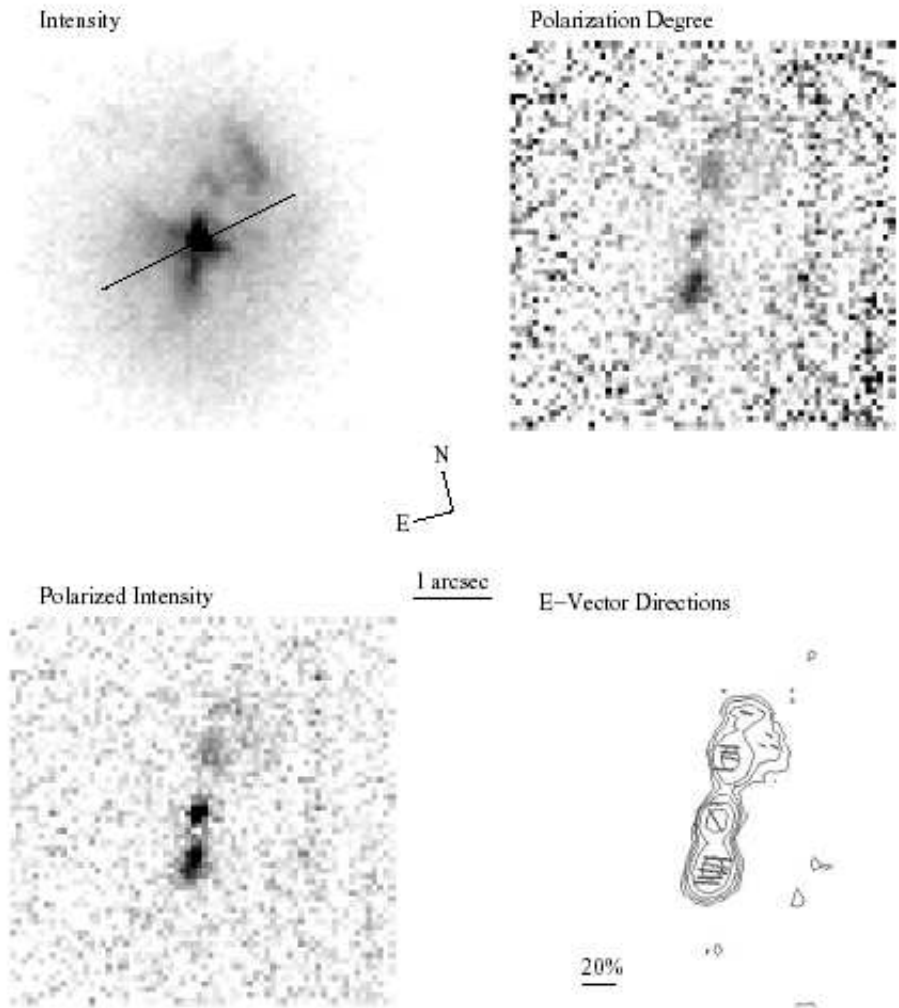}
\caption{HST/NICMOS 2.0$\mu$m intensity and polarization images for the nuclear regions of the powerful
radio galaxy Cygnus A (see Tadhunter et al. 2000). The line segment in the intensity image
indicates the direction of the radio axis.}
\end{figure}

\subsection{Scattering outflows in the NLR of Cygnus A}

A further interesting feature of the near-IR intensity images of Cygnus A is that they reveal the
cones to be hollowed out close to the nucleus (see Figure 3, top left), presumably as a consequence of outflows
driven by the powerful AGN activity (Tadhunter et al. 1999). 
The question then arises as to whether there is any
kinematic evidence for the outflows that are hollowing out the cones.

Outflows are notoriously difficult to detect in the extended regions of distant active
galaxies, because of uncertainties about the true systemic velocities of the host galaxies,
and the likely combination of gravitational (inflow, rotational) and AGN-induced
(outflow) motions.   As part of a detailed study of Cygnus A designed to measure the
mass of the central black hole, we have used the [OIII]$\lambda$5007 line to map the emission line
kinematics close to the AGN at 0.1 arcsec (0.1 kpc) resolution using STIS/HST (Tadhunter et al. 2003). 
The
steep velocity gradient across the nucleus revealed by these observations is consistent with
rotational motion about a (2.5$\pm0.5$)$\times$10$^9$ M$_{\odot}$ black hole. 
However, we also detect a
component 0.2 arcsec from the nucleus in the centre of the cone that is redshifted by $\sim$400 km s$^{-1}$
relative to the rest frame of the host galaxy. This component cannot be reconciled with
rotational motion associated with either the central disk or the kpc-scale dust lane. It is
also difficult to explain the redshift of this component in terms of direct [OIII]  emission 
from the AGN-induced outflow, since the NW cone is tilted towards our line of sight.

Clues to the nature of the strongly redshifted component are provided by
the spectropolarimetry observations reported by van Bemmel at this conference (see also
van Bemmel et al. 2003). In an aperture that encompasses the entire nuclear region of
Cygnus A, van Bemmel et al. detect an [OIII] feature in the polarized intensity spectrum
that is redshifted by the same amount ($\sim$350 ­- 450 km s$^{-1}$) as the redshifted
component detected along
the radio axis in our STIS/HST data. Therefore it is likely that this latter component represents
[OIII] emission from an inner narrow line region that has been scattered by
outflowing dust in the NW cone. Since a scattering outflow will always produce a redshifted
feature, this explanation avoids the geometrical problem of attempting to explain the redshifted feature
in terms of direct [OIII] emission from an outflow in the NW cone.

Clearly, deep spectropolarimetry observations of the narrow line emission provide a
useful tool for investigating the significance of activity-induced outflows in powerful
radio galaxies. It will be important to extend such observations to larger 
samples of radio galaxies in the
future.

\section{Conclusions}

Over the last 20 years optical/IR polarimetry has emerged as 
a key technique for investigating several aspects of the radio galaxy
phenomenon. The diversity of the science
made possible by polarimetric studies is impressive, ranging from 
testing the unified schemes and 
determining the structure of the 
near-nuclear regions, to investigating the nature of the host galaxies of 
some of the most distant known active galaxies.
 
Polarimetry is likely to remain an important technique in the future, with 
particular emphasis on using polarimetric observations to probe the 
sub-kpc structures, quantify the impact of the activity on the host 
galaxies, and assess the degree of AGN pollution of the host galaxy 
colours.

Although this review has concentrated on radio galaxies, many of 
the conclusions also 
apply to other classes of AGN. For example, the AGN pollution of the host 
galaxy colours, which is becoming well-quantified for radio galaxies, is 
likely to be serious (but so far ignored!) issue for the host 
galaxies of quasars and other types of distant AGN.

\newpage\noindent
{\bf Appendix I: Optical/UV polarization studies of radio galaxies}
\vglue 0.3cm\noindent
{\small {\bf Table 1.} Compilation of references for optical/UV polarimetry observations of radio galaxies.
The
first column gives the references in chronological order. The second column gives the
type of polarimetry observation (Key: Sp -- spectropolarimetry; Im -- imaging
polarimetry (with polarization map); Ap -- aperture polarimetry at single wavelength;
Ap(Im) -- aperture polarimetry measurement derived from imaging polarimetry data). The
third column gives the number of objects included  in the study. The final column gives the redshift range covered by the
study.}
\begin{table}[!h]
%\smallskip

{\small 

\begin{center}
\begin{tabular}{llll}
\tableline
\noalign{\smallskip}
Reference &Type of study &Number of objects &Redshift range \\
\noalign{\smallskip}
\tableline
\noalign{\smallskip}
Antonucci (1982) &Sp &10 &$z < 0.256$ \\
Antonucci (1984) &Sp &31 &$z < 0.256$ \\
di Serego Alighieri et al. (1988) &Ap(Im) &1(PKS2152-69) &$z=0.0283$ \\
di Serego Alighieri et al. (1989) &Ap(Im) &2 &$z=0.766$ \& $1.132$ \\
Tadhunter et al. (1990) &Im &1(Cygnus A) &$z=0.0560$ \\
Scarrott et al. (1990) &Im &1(3C368) &$1.132$ \\
Januzzi \& Elston (1991) &Ap(Im) &1(3C265) &$z=0.81$ \\
Impey et al. (1991) &Ap &20 &$z < 0.549$ \\
Goodrich \& Cohen (1992) &Sp &1(3C109) &$z=0.3066$ \\
Tadhunter et al. (1992) &Ap(Im) &12 &$0.2 < z < 0.85$ \\
Jackson \& Tadhunter (1993) &Sp &1(Cygnus A) &$z=0.056$ \\
di Serego Alighieri et al. (1993) &Ap(Im) &6 &$0.380 < z < 1.206$ \\
Draper et al. (1993) &Im &4 &$0.041 < z < 0.22$ \\
Cimatti et al. (1993) &Ap(Im),Sp &42 &$0.096 < z < 1.206$ \\
di Serego Alighieri et al. (1994) &Sp &3 &$0.766 < z < 0.818$ \\
Tadhunter et al. (1994) &Ap(Im) &1(PKS1934-63) &$z = 0.183$ \\
Cimatti \& di Serego Alighieri (1995) &Im,Ap(Im) &8 &$0.096 < z < 1.206$ \\
Shaw et al. (1995) &Im(Ap) &4 &$0.3 < z < 0.7$ \\
Tran et al. Cohen (1995) &Sp &1(3C234) &$z=0.1848$ \\
Dey \& Spinrad (1996) &Sp &1(3C265) &$z=0.81$ \\
Cimatti et al. (1996) &Sp &1(3C324) &$z=1.2$ \\
Dey et al. (1996) &Sp &1(3C256) &$z=1.824$ \\
Tadhunter et al. (1996) &Sp &1(3C321) &$z=0.096$ \\
Cimatti et al. (1997) &Sp &2 &$z=1.08$ \& $1.35$ \\
di Serego Alighieri et al. (1997) &Sp &9 &$0.086 < z < 0.258$ \\
Ogle et al. (1997) &Sp,Im &1(Cygnus A) &$z=0.056$ \\
Cohen et al. (1997) &Sp,Im,Ap &1(PKS0116+082) &$z = 0.594$ \\
Dey et al. (1997) &Sp &1(4C41.17) &$z = 3.8$ \\
Cimatti et al. (1998) &Sp &2 &$z=2.482$ \& $2.366$ \\
Tran et al. (1998) &Sp,Im &4 &$0.75 < z < 1.2$ \\
Corbett et al. (1998) &Sp &5 &$0.057 < z < 0.286$ \\
Brotherton et al. (1998) &Sp &1(3C68.1) &$z=1.228$ \\
Hurt et al. (1999) &Im(UV) &8 &$0.056 < z < 0.58$ \\
Cohen et al. (1999) &Sp,Im &13 &$0.056 < z < 0.185$ \\
Corbett et al. (2000) &Sp &4 &$0.057 < z < 0.371$ \\
Vernet et al. (2001) &Sp &9 &$2.26 < z < 3.56$ \\
Kishimoto et al. (2001) &Sp,Im(UV) &3 &$0.096 < z < 0.185$ \\
Tadhunter et al. (2002) &Ap(Im) &19 &$0.15 < z < 0.7$ \\
van Bemmel et al. (2003) &Sp &1(Cygnus A) &$z=0.056$ \\
\noalign{\smallskip}
\tableline

\end{tabular}

\end{center}
}
\end{table}

\end{document}